\documentclass[12pt]{aastex}

\begin{document}


\title{The Influence of Uncertainties in the $^{15}$O($\alpha,\gamma)^{19}$Ne Reaction Rate on Models of Type I X-Ray Bursts}


\author{Barry Davids}
\affil{TRIUMF, Vancouver, BC, Canada}

\author{Richard H. Cyburt}
\affil{Joint Institute for Nuclear Astrophysics and National Superconducting Cyclotron Laboratory, Michigan State University, East Lansing, MI}

\author{Jordi Jos\'e}
\affil{Dept.\ Fisica i Enginyeria Nuclear, Universitat Polit\`ecnica de Catalunya and Institut d'Estudis Espacials de Catalunya, Barcelona, Spain}

\and

\author{Subramanian Mythili}
\affil{Physics Department, George Mason University, Fairfax, VA}




\begin{abstract}
We present a Monte Carlo calculation of the astrophysical rate of the $^{15}$O($\alpha,\gamma)^{19}$Ne reaction based on an evaluation of published experimental data. By considering the likelihood distributions of individual resonance parameters derived from measurements, estimates of upper and lower limits on the reaction rate at the 99.73\% confidence level are derived in addition to the recommended, median value. These three reaction rates are used as input for three separate calculations of Type I x-ray bursts using spherically symmetric, hydrodynamic simulations of an accreting neutron star. In this way the influence of the $^{15}$O($\alpha,\gamma)^{19}$Ne reaction rate on the peak luminosity, recurrence time, and associated nucleosynthesis in models of Type I x-ray bursts is studied. Contrary to previous findings, no substantial effect on any of these quantities is observed in a sequence of four bursts when varying the reaction rate between its lower and upper limits. Rather, the differences in these quantities are comparable to the burst-to-burst variations with a fixed reaction rate, indicating that uncertainties in the $^{15}$O($\alpha,\gamma)^{19}$Ne reaction rate do not strongly affect the predictions of this Type I x-ray burst model. 
\end{abstract}





\section{Introduction}

Type I x-ray bursts are thought to arise from thermonuclear runaways occurring in a thin layer on the surfaces of neutron stars accreting hydrogen- and helium-rich matter from companion stars in close binary systems \citep{woosley76,joss77}. Accreting neutron stars in binary systems stably fuse hydrogen into helium via the temperature-independent hot CNO cycles at high accretion rates. At low accretion rates they exhibit x-ray bursts due to the explosive burning of H or He of typical duration 10-100 s during which approximately $10^{39-40}$ erg is liberated \citep{cumming04}. If the accreted material is rich in H, the $rp$ process occurs after the burst ignites, potentially leading to the synthesis of nuclei up to $A \approx 100$ \citep{schatz01} or beyond \citep{koike04,elomaa09}. The light nuclei that participate in the CNO cycles and the heavier nuclei of the $rp$ process are linked by the breakout reactions $^{15}$O($\alpha,\gamma)^{19}$Ne and $^{18}$Ne($\alpha,p)^{21}$Na \citep{wallace81,wiescher99}. The latter reaction becomes important at higher temperatures than does the former; hence $^{15}$O($\alpha,\gamma)^{19}$Ne has a larger effect on the hydrogen mass fraction at the time of burst ignition than does $^{18}$Ne($\alpha,p)^{21}$Na, thereby more strongly influencing the dynamics of the burst. For this reason, the effect of this rate on x-ray burst models has been the subject of several investigations \citep{fisker06,cooper06,fisker07}.

While the $^{15}$O($\alpha,\gamma)^{19}$Ne reaction rate has not been measured directly, intense and steady experimental efforts have enabled measurements of the decay properties of the states in $^{19}$Ne relevant to the reaction rate at x-ray burst temperatures. These states are just above the $\alpha$ threshold which lies at an excitation energy of 3.53 MeV. As they lie below the proton and neutron thresholds, the states of interest decay only by $\alpha$ or $\gamma$ ray emission. Hence the resonant rate of $^{15}$O($\alpha,\gamma)^{19}$Ne can be calculated from the radiative and $\alpha$ widths, or equivalently the lifetimes and $\alpha$-decay branching ratios of the relevant states.

In this paper we assess the experimental data on the decay properties of states just above the $\alpha$ emission threshold in $^{19}$Ne and use a Monte Carlo technique to evaluate the reaction rate. With the results of this analysis upper and lower limits on the reaction rate at the 99.73\% confidence level are derived in addition to the recommended, median value. These three reaction rates are used as input for three separate calculations of Type I x-ray bursts using spherically symmetric, hydrodynamic simulations of an accreting neutron star. In this way the influence of the $^{15}$O($\alpha,\gamma)^{19}$Ne reaction rate on the peak luminosity, recurrence time, and associated nucleosynthesis of Type I x-ray bursts is studied and compared with previous findings.

\section{Decay Properties of $^{19}$Ne States and the Resonant Rate of $^{15}$O($\alpha,\gamma)^{19}$Ne}

The thermonuclear rate of the $^{15}$O($\alpha,\gamma)^{19}$Ne reaction at temperatures below 2 GK, the relevant range for Type I x-ray bursts, is dominated by the contributions of resonances corresponding to states in $^{19}$Ne with excitation energies between 4 and 5 MeV. These states lie above the $\alpha$ decay threshold but below the thresholds for neutron or proton decay. Hence they decay only by $\alpha$ and $\gamma$ emission and the total decay width $\Gamma$ is the sum of these partial decay widths, i.e., $\Gamma=\Gamma_\alpha+\Gamma_\gamma$. As these resonances are narrow compared to the spacing between adjacent levels with the same spin and parity, the contributions of the different resonances add incoherently and can be measured separately. Even in the absence of a direct measurement, the strength of a resonance can be inferred from measurements of the radiative and $\alpha$ widths of a state. Equivalently, measurements of the mean lifetime of a state $\tau$ and its $\alpha$-decay branching ratio $B_\alpha\equiv\Gamma_\alpha/\Gamma$ suffice to determine the strength of a resonance. The resonance strength $\omega\gamma$ of a $^{19}$Ne state with angular momentum $J$ formed by the capture of an $\alpha$ particle by $^{15}$O is given by \begin{equation}
\omega\gamma = \frac{2J+1}{(2J_\alpha+1)(2J_{^{15}O}+1)}\frac{\Gamma_\alpha\Gamma_\gamma}{\Gamma}= \frac{2J+1}{2}\frac{(B_\alpha\Gamma)(1-B_\alpha)\Gamma}{\Gamma}=\frac{2J+1}{2}B_\alpha(1-B_\alpha)\Gamma.
\end{equation} Taking account of the fact that the mean lifetime and total decay width are related by $\Gamma=\hbar/\tau$, this implies that \begin{equation}
\omega\gamma=\frac{(2J+1)B_\alpha(1-B_\alpha)\hbar}{2\tau}.
\end{equation}

The results of a number of $B_\alpha$ measurements of states in $^{19}$Ne have been published over the last twenty years \citep{magnus90,laird02,davids03a,davids03,rehm03,visser04,tan07,tan09}. In addition, the lifetimes of these states have been measured by two independent groups over the past five years \citep{tan05,kanungo06,mythili08}. Here we critically evaluate the published measurements.

\subsection{Mean Lifetime Data}

All published lifetime measurements for states with excitation energies up to 4.602 MeV in $^{19}$Ne are shown in Table \ref{tablifetime}. At least two measurements of the lifetimes of all the known states with excitation energies between 1536 keV and 4602 keV have been published and agree. This agreement gives us a measure of assurance that the values are reliable, though apart from the state at 4.03 MeV, the only Doppler shifted $\gamma$-ray spectra for states above the $\alpha$ threshold that have been published in the literature appear in \citet{mythili08}. For the states lying above 4602 keV, no published lifetime data exist, likely because the states decay predominantly by $\alpha$ decay and therefore were not observed in Doppler shift attenuation measurements. We use analog state information to deduce information on the radiative widths of the 4.71 and 5.09 MeV states. Although there are known states above 5.092 MeV and below the proton threshold, they all have $\alpha$ decay branching ratios consistent with 1 \citep{rehm03} and their high excitation energies render them unimportant contributors to the thermally averaged reaction rate at Type I X-ray burst temperatures. Hence we make no attempt to estimate the lifetimes of these states. 

\begin{table}
\caption{$^{19}$Ne level mean lifetime measurements and $1\sigma$ uncertainties in fs. Statistical and systematic errors are listed separately in the last column.\label{tablifetime}}
\begin{tabular}{cccc}
\tableline\tableline
Level Energy (MeV) & \cite{tan05} & \cite{kanungo06} & \cite{mythili08}\\
\tableline
1.54 & $16 \pm 4$ & & $19.1^{+0.7}_{-0.6}\pm1.1$\\
4.03 & $13^{+9}_{-6}$ & $11^{+4}_{-3}$ & $6.9\pm1.5\pm0.7$\\
4.14 & $18^{+2}_{-3}$ & & $14.0^{+4.2}_{-4.0}\pm1.2$\\
4.20 & $43^{+12}_{-9}$ & & $38^{+20}_{-10}\pm2$\\
4.38 & $5^{+3}_{-2}$ & & $2.9\pm1.4\pm0.6$\\
4.55 & $15^{+11}_{-5}$& & $18.7^{+3.0}_{-2.6}\pm2.2$\\
4.60 & $7^{+5}_{-4}$ & & $7.6^{+2.1}_{-2.0}\pm0.9$\\
\tableline
\end{tabular}
\end{table}

\subsection{Spin and Parity Assignments}

In the relevant excitation energy range, the spins and parities of most of the states are well known. The exceptions are two of the negative parity states, the 4144 keV state with a tentatively assigned spin of 9/2 and the 4200 keV state tentatively assigned to be spin 7/2. \citet{garrett72} initially discussed the 4144 keV state as a 7/2$^-$ state and the 4200 keV state as a 9/2$^-$ state in the context of the known rotational bands in $^{19}$F and $^{19}$Ne. Based on their DWBA analyses \citet{garrett72} suggested swapping the spin assignments of these two states, a suggestion tentatively adopted by evaluators \citep{tilley95}. 

We compare the spins of the $^{19}$Ne states with the spins of the presumed isobaric analog states in $^{19}$F by examining the reduced transition probabilities and the branching ratios of putative analog states. The isoscalar component dominates strong $E2$ transitions. The 4033 keV state in $^{19}$F undergoes a strong $E2$ transition. Depending on the spin, either the 4144 keV state in $^{19}$Ne or the 4200 keV state undergoes an $E2$ transition. Both the 3999 and 4033 keV states in $^{19}$F decay to the 5/2$^-$ state at 1346 keV. We calculate the reduced transition probabilities of the $^{19}$Ne states using the lifetimes of \citet{mythili08} and transition energies of \citet{tilley95} and \citet{tan05}. In $^{19}$F, the 3999 keV state decays to the 1346 keV state via an $M1$ transition while the 4033 keV state undergoes an $E2$ transition. These transition strengths are tabulated in the first row of Table \ref{be2table1}. The 4144 keV and 4200 keV states in $^{19}$Ne decay to the 5/2$^-$ state at 1508 keV. If we accept the tentative spin assignments of 9/2$^-$ and 7/2$^-$ respectively, the 4200 keV state undergoes an $M1$ transition and the 4144 keV state an $E2$ transition. The resulting reduced transition probabilities appear in the second row of Table \ref{be2table1}. The reduced transition probabilities calculated with the alternate spin assignments are shown in the third row of Table\ \ref{be2table1}, where we see improved agreement with the $^{19}$F $B(E2)$ value.

\begin{table}[h]
\begin{tabular}{cccccccccc}
\tableline\tableline
Nucleus &  $J^{\pi}$ & Energy Level &$B(M1)$&Nucleus &  $J^{\pi}$ & Energy Level &$B(E2)$ \\
&&(keV)&(MeV fm$^{3}$)&&&(keV)&(MeV fm$^{5}$)\\
\tableline
$^{19}$F & $\frac{7}{2}^{-}$ &3999&0.0017$^{+0.0010}_{-0.0005}$&$^{19}$F & $\frac{9}{2}^{-}$ &4033&90$\pm$20 \ \\
&&\\
$^{19}$Ne & ($\frac{7}{2})^{-}$ &4200&0.0008$^{+0.0003}_{-0.0003}$&$^{19}$Ne & ($\frac{9}{2})^{-}$ &4144&460$^{+180}_{-100}$\ \\
&&\\
$^{19}$Ne & ($\frac{7}{2})^{-}$ &4144&0.0024$^{+0.0010}_{-0.0009}$&$^{19}$Ne & ($\frac{9}{2})^{-}$ &4200&150$^{+60}_{-50}$\\
&&\\
\hline
\end{tabular}
\caption{Reduced $M1$ and $E2$ transition probabilities for two isobaric analog states in $^{19}$F and $^{19}$Ne. The first row shows the transition strengths for two states in $^{19}$F. The second row contains the $^{19}$Ne transition strengths with the tentatively adopted spin assignments of \citet{tilley95} for the $^{19}$Ne analog states and the third row gives the transition strengths with the alternate assignment of the $^{19}$Ne spins. }
\label{be2table1} 
\end{table}

Although the reduced transition probabilities suggest that perhaps the spins should be swapped, the measured $\gamma$ decay branching ratios support the tentative spin assignments of \cite{tilley95} for the 4144 keV and 4200 keV states in $^{19}$Ne. The 9/2$^-$ state at 4033 keV in $^{19}$F decays exclusively to the 1346 keV state;  the 4144 keV state in $^{19}$Ne decays exclusively to the 1508 keV, suggesting that the tentative spin assignment is justified. The 4200 keV state in $^{19}$Ne and the 3999 keV state in $^{19}$F have comparable branching ratios of 80\% and  70\% to the 1508 keV state in $^{19}$Ne and the 1346 keV state in $^{19}$F respectively. It is necessary that the spins of the 4144 and 4200 keV states be measured experimentally in order to precisely constrain their contributions to the reaction rate, though this is not the largest uncertainty; for now we adopt the tentative spin assignments of \citet{tilley95}.

\subsection{$\alpha$ Decay Branching Ratio Data}

Independent measurements of the $\alpha$ decay branching ratio have yielded good agreement for the high lying states, but measurements of the low lying states have been controversial.  All published measurements of which we are aware are shown in Table \ref{tabbalpha1} and Table \ref{tabbalpha2}. The first measurements were reported by \citet{magnus90} and are shown in Table \ref{tabbalpha1}. The lowest energy state whose $\alpha$ decay branching ratio was reported in this measurement was the 4.38 MeV state, for which $B_\alpha=0.044\pm0.032$ was given. Examination of the timing spectrum for this state shown in \citet{magnus90} reveals no peak and leads us to conclude that, as one would expect {\it a priori}, a 1.4$\sigma$ excess over background such as this does not warrant a detection claim; these data can only be used to constrain $B_\alpha$ from above. In contrast, all the other $B_\alpha$ measurements reported in \citet{magnus90} appear to be based on statistically significant excesses observed over well-determined backgrounds.

\begin{table}
\caption{$^{19}$Ne level $\alpha$ decay branching ratio $B_\alpha$ measurements and $1\sigma$ uncertainties. Upper limits are cited at the 90\% confidence level.\label{tabbalpha1}}
\begin{tabular}{cccc}
\tableline\tableline
Level Energy (MeV) & \cite{magnus90} & \cite{laird02} & \cite{davids03}\\
\tableline
4.03 &  &  & $<4.3\times10^{-4}$\\
4.38 & $0.044\pm0.032$ & & $<3.9\times10^{-3}$\\
4.55 &$ 0.07\pm0.03$ & & $0.16\pm0.04$\\
4.60 & $0.25\pm0.04$ & $0.32\pm0.03$ & $0.32\pm0.04$\\
4.71 & $ 0.82\pm0.15$ & & $0.85\pm0.04$\\
5.09 & $ 0.90\pm0.09$ & & $0.90\pm0.06$\\
\tableline
\end{tabular}
\end{table}

\citet{laird02} measured $B_\alpha$ for a single state and found a value consistent with that of \citet{magnus90}.

The measurement reported in \citet{davids03} was the most sensitive made to date of the $\alpha$ decay of the 4.03 MeV state owing to the small, well-measured background and the selectivity of the $p(^{21}$Ne,$t)^{19}$Ne reaction used to populate it. Contrary to the claim made in \citet{tan09}, net $\alpha$ decay events above the estimated background were observed at the location of the 4.03 MeV level in the measurement of \citet{davids03}. However this small excess was not deemed statistically significant and the measurement was reported as an upper limit. No excess above the background was observed at the location of the 4.38 MeV state and therefore an upper limit on $B_\alpha$ was also reported for this state. We note that this 90\% confidence level upper limit on $B_\alpha$ for the 4.38 MeV state is more than a factor of ten smaller than the central value of the statistically insignificant detection of $\alpha$ decay reported from this state in \citet{magnus90}, illustrating the peril of relying on gaussian representations of the likelihood distributions of such measurements. The $B_\alpha$ values for the other states reported in \citet{davids03} are consistent with those of \cite{magnus90}.

\citet{rehm03} report $B_\alpha$ measurements of three states in $^{19}$Ne but the excitation energy resolution was around 220 keV and was insufficient to resolve adjacent levels, rendering the fitting used to deduce the two $B_\alpha$ central values and associated uncertainties less than wholly satisfying. Nevertheless there was no background around the 4.03 MeV level, permitting a firm upper limit on $B_\alpha$ for this state to be established. 


Problems with the electronics used in the measurement of \citet{visser04} rendered the normalization of the B$_\alpha$ values reported there questionable. In particular, the measurements were normalized to the coincidence efficiency determined based on B$_\alpha$ measurements of the 5351 keV level. The measurement yielded different values of B$_\alpha$ for this state, all of which differed substantially from the accepted value of 1, depending on the different electronics gating configurations used in the measurement. Out of concern that the normalization is not completely understood, we have avoided using these data. 

Although the relevant states were populated and separated with good energy resolution in the measurement of \citet{tan07}, the large background was not persuasively modelled for the states below 4.55 MeV from which $\alpha$ decay was putatively detected. Since the backgrounds were not well determined in the 4.03, 4.14/4.20, and 4.38 MeV state $\alpha$ decay energy spectra shown in Figure 3 of \citet{tan07}, the excesses claimed to be observed above them are not quantitatively reliable. 
Inspection of the timing spectra for these putative $\alpha$ decay detections from states below 4.55 MeV in $^{19}$Ne, shown in Figure 6 of \citet{tan09}, reveals no convincing evidence of an $\alpha$ decay peak from the 4.03, 4.14, 4.20, or 4.38 MeV states. In contrast the B$_\alpha$ values for states lying at and above 4.55 MeV appear reliable as far as the background determination and signal to noise ratio are concerned, and they agree with the measurements of \citet{magnus90}.

\begin{table}
\caption{$^{19}$Ne level $\alpha$ decay branching ratio $B_\alpha$ measurements and $1\sigma$ uncertainties. Upper limits are cited at the 90\% confidence level.\label{tabbalpha2}}
\begin{tabular}{cccc}
\tableline\tableline
Level Energy (MeV) & \cite{rehm03} & \cite{visser04} & \cite{tan09}\\
\tableline
4.03 & $<6\times10^{-4}$ &  & $2.9\pm2.1\times10^{-4}$\\
4.14 and 4.20 &  & & $1.2\pm0.5\times10^{-3}$\\
4.38 & $ 0.016\pm0.005$ & & $1.2\pm0.3\times10^{-3}$\\
4.55 & &  $0.06\pm0.04$ & $0.07\pm0.02$\\
4.60 & & $0.208\pm0.026$ & $0.26\pm0.03$\\
4.71 & & $0.69^{+0.11}_{-0.14}$ & $0.80\pm0.15$\\
5.09 & $ 0.8\pm0.1$  &$0.75^{+0.06}_{-0.07}$& $0.87\pm0.03$\\
\tableline
\end{tabular}
\end{table}

\section{Monte Carlo Reaction Rate Calculation}

We report here the results of a million event Monte Carlo simulation of the thermally averaged rate of the $^{15}$O($\alpha,\gamma)^{19}$Ne per particle pair as a function of temperature. This reaction rate is the incoherent sum of the nonresonant rate, as calculated by \citet{dufour00}, and the contributions due to the $^{19}$Ne states at 4.03, 4.14, 4.20, 4.38, 4.55, 4.60, 4.71, and 5.09 MeV. The nonresonant contribution is negligible above 0.2 GK. While the nonresonant rate is fixed at a central value due to its totally insignificant impact on the error budget, the resonant contributions are calculated by drawing randomly from the likelihood distributions of the resonance parameters. The energies and spins of the levels adopted here are shown in Table \ref{tablevels}. The contribution due to each resonance was calculated as follows. The energy and 1$\sigma$ uncertainty of each state was taken from the compilation of \citet{tilley95} or calculated by using a weighted average of this and the value from \citet{tan05}, including a discrepancy error to account for potentially discrepant data as described in \citet{cyburt01}. The likelihood distributions for the level energies are taken to be gaussians characterized by the means and standard deviations given in Table \ref{tablevels}.

\begin{table}
\caption{$^{19}$Ne level angular momenta and energies with $1\sigma$ uncertainties adopted in this work.\label{tablevels}}
\begin{tabular}{ccc}
\tableline\tableline
Level Energy (keV) & J$^\pi$ & Reference(s) \\
\tableline
4034.3(9) & $3/2^+$ & \citet{tilley95,tan05}\\
4143.4(8) & ($9/2)^-$ & \citet{tilley95,tan05}\\
4199.7(16) & ($7/2)^-$ & \citet{tilley95,tan05}\\
4377.9(7) & $7/2^+$ & \citet{tilley95,tan05}\\
4547.8(10) & $3/2^-$ & \citet{tilley95,tan05}\\
4601.7(9) & $5/2+$ & \citet{tilley95,tan05}\\
4712(10) & $5/2^-$ & \citet{tilley95}\\
5092(6) & $5/2^+$ & \citet{tilley95}\\
\tableline
\end{tabular}
\end{table}

\begin{figure}
\includegraphics[width=\linewidth]{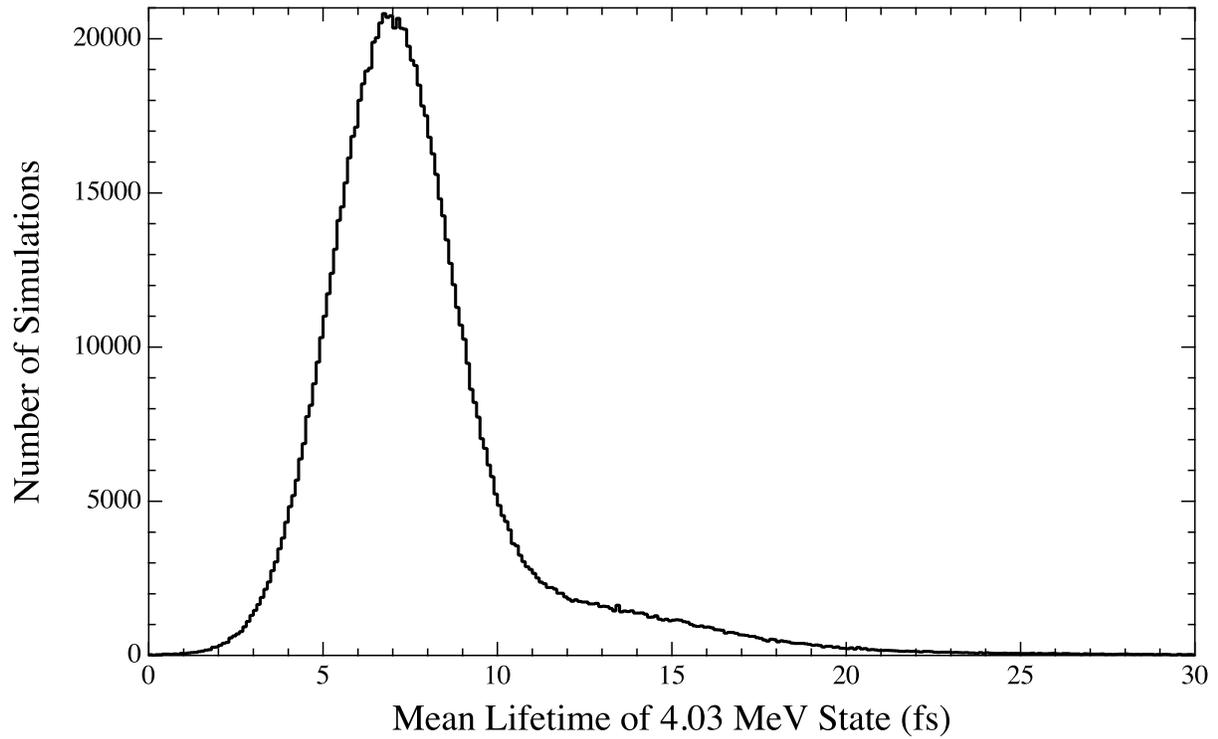}
\caption{Likelihood distribution of the mean lifetime of the 4.03 MeV state in $^{19}$Ne based on the measurements of Tan \emph{et al.} 2005, Kanungo \emph{et al.} 2006, and Mythili \emph{et al.} 2008. The weight assigned to each measurement is proportional to the inverse of its variance.}
\label{tau403}
\end{figure}

\begin{figure}
\includegraphics[width=\linewidth]{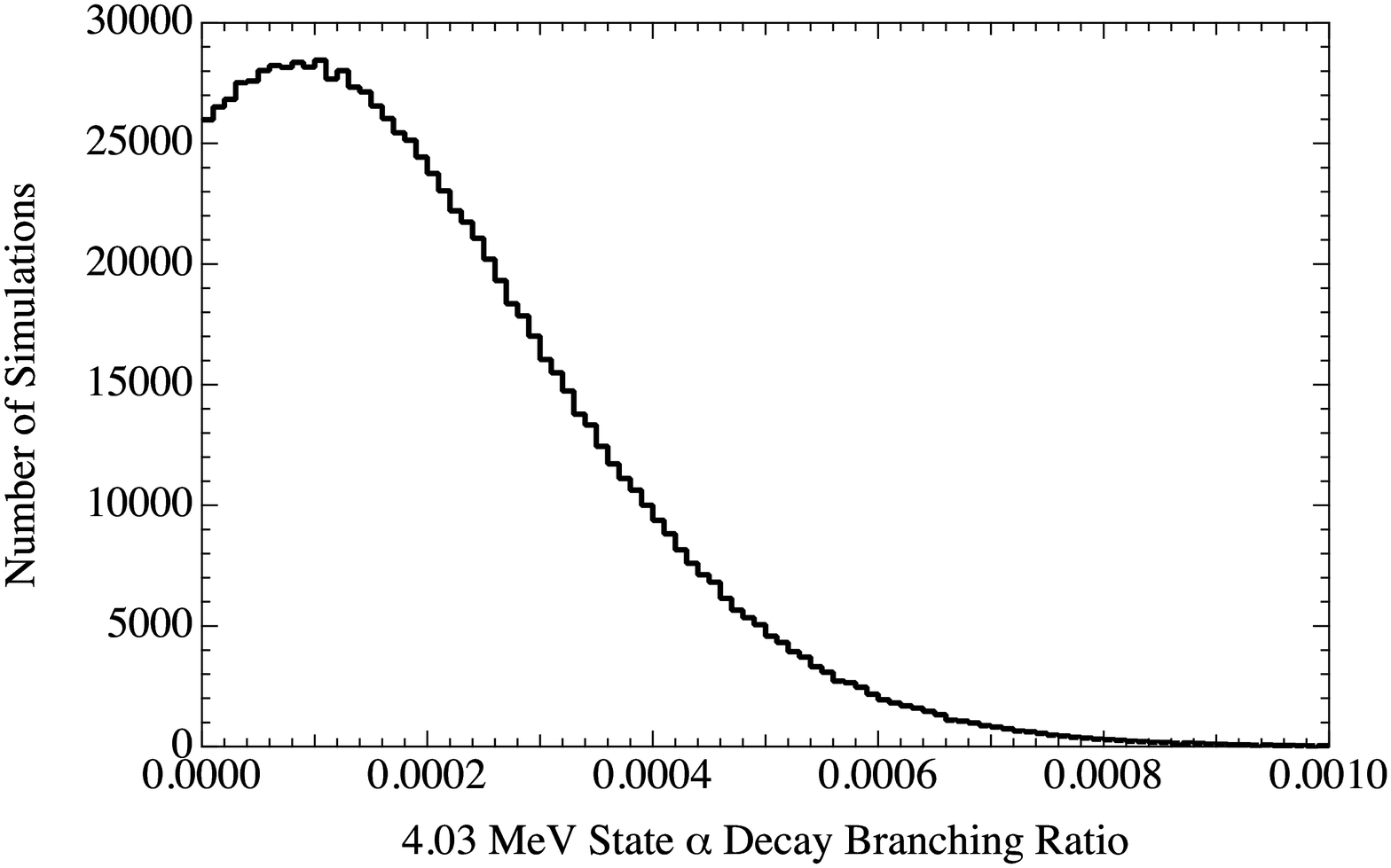}
\caption{Likelihood distribution of the $\alpha$ decay branching ratio of the 4.03 MeV state in $^{19}$Ne based on the measurement of Davids \emph{et al.} 2003.}
\label{ba403}
\end{figure}

The lifetime of the 4.03 MeV state is drawn from the likelihood distributions for this quantity reported by \citet{tan05}, \citet{kanungo06}, and \citet{mythili08}, with the weight of each measurement assigned according to the inverse of its variance. The resulting likelihood distribution realized in the one million event Monte Carlo simulation is shown in Figure \ref{tau403}. For B$_\alpha$ of this state we use the likelihood distribution derived from the measurement of \citet{davids03}, shown in Figure \ref{ba403}.

Prior to the measurement of \citet{davids03}, the best information on $\Gamma_\alpha$ for the 4.03 MeV state came from measurements of $\alpha$ transfer to the $^{19}$F analog state \citep{mao95}. The uncertainties in using isospin symmetry to deduce reduced alpha widths of states near threshold from their analog states are fairly well quantified \citep{fortune10}, but systematic uncertainties due to the reaction mechanism remain. Fortunately there are reliable experimental data on $B_\alpha$ for the 4.03 MeV state and we need not rely on the analog state. Moreover the most likely value of $\Gamma_\alpha$ for the 4.03 MeV state derived from the Monte Carlo simulation is 8 $\mu$eV, perfectly consistent with the value of 8.8(14) $\mu$eV given (for q~=~9) in \citet{mao95}.

\begin{figure}
\includegraphics[width=\linewidth]{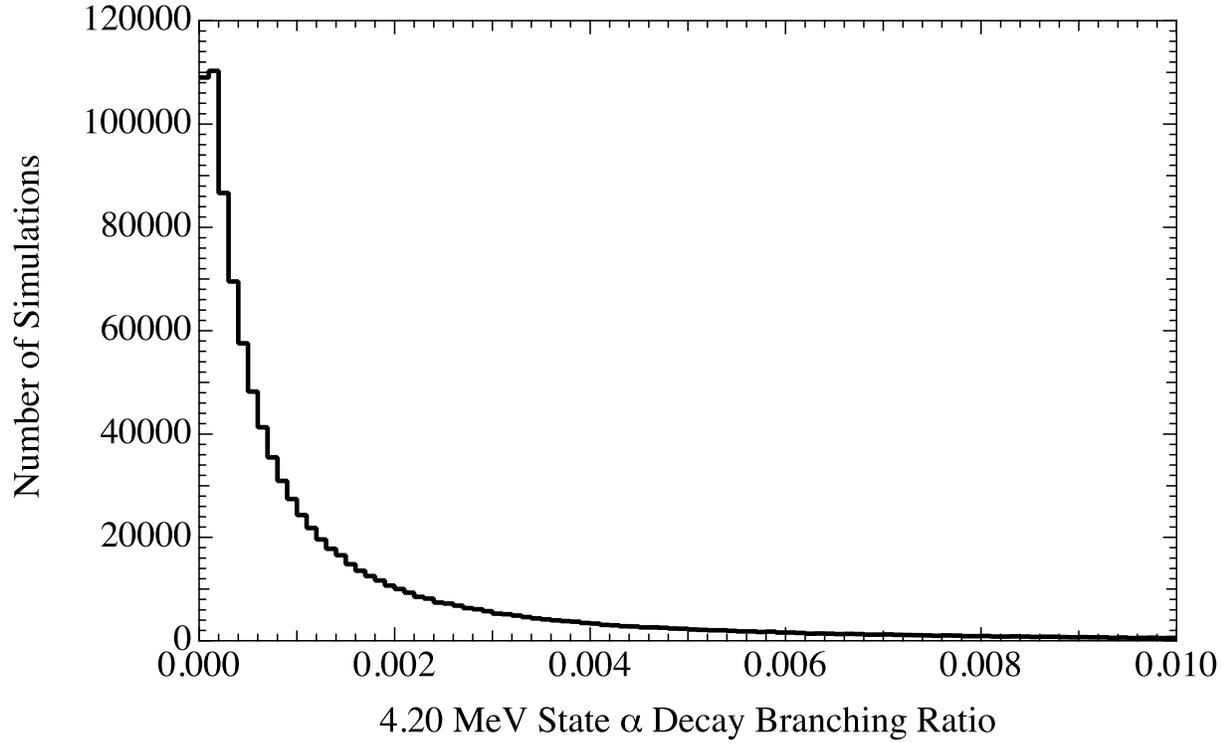}
\caption{Likelihood distribution of the $\alpha$ decay branching ratio of the 4.20 MeV state in $^{19}$Ne. We compute this using R-matrix formulae assuming the reduced $\alpha$ width is given by a log normal distribution peaked at 0.005 and having a standard deviation of its logarithm of 1.5.}
\label{ba420}
\end{figure}

The mean lifetimes of the 4.14 and 4.20 MeV states are drawn from the measurements of \citet{tan05} and \citet{mythili08} as described above. As we are not convinced of the reliability of the single B$_\alpha$ measurement for these two states reported in the literature, we calculate the $\alpha$ widths of these states according to
\begin{equation}
\Gamma_\alpha=\frac{3 \hbar^2}{\mu a^2} P \theta_\alpha^2,
\end{equation}
where $\mu$ is the reduced mass of the $^{15}$O$ + \alpha$ system, $a$ the channel radius, $P$ the Coulomb penetrability, and $\theta_\alpha^2$ the reduced $\alpha$ width \citep{tw52}. Despite some efforts to determine $\theta_\alpha^2$ for the $^{19}$F analog states \citep{oliveira96,mao96}, in our judgement no reliable experimental determinations of these reduced widths are available in the literature, so we draw them from distributions of our own construction. We conjecture that these reduced widths can be adequately represented by a log normal distribution having a mode of $\theta_\alpha^2=0.005$ and standard deviation of its logarithm of 1.5. In a survey of  $\theta_\alpha^2$ values of unbound states in the $\alpha$ nuclei $^{24}$Mg, $^{28}$Si, $^{32}$S, $^{36}$Ar, and $^{40}$Ca, \citet{longland10} find a mean value of 0.01. For comparison, the $\theta_\alpha^2$ value given in \citet{oliveira96} (without any uncertainty) is 0.135 for both states. We chose a channel radius of $a=5.8$ fm, corresponding to $r_0=1.45$ fm; we investigated different channel radii of 5.6 and 6.0 fm and found that the resulting penetrabilities changed by $\sim50$\%, which is totally insignificant compared with the uncertainty due to the reduced width. The likelihood distribution for the B$_\alpha$ value of the 4.20 MeV state resulting from this calculation is shown in Figure \ref{ba420}.

\begin{figure}
\includegraphics[width=\linewidth]{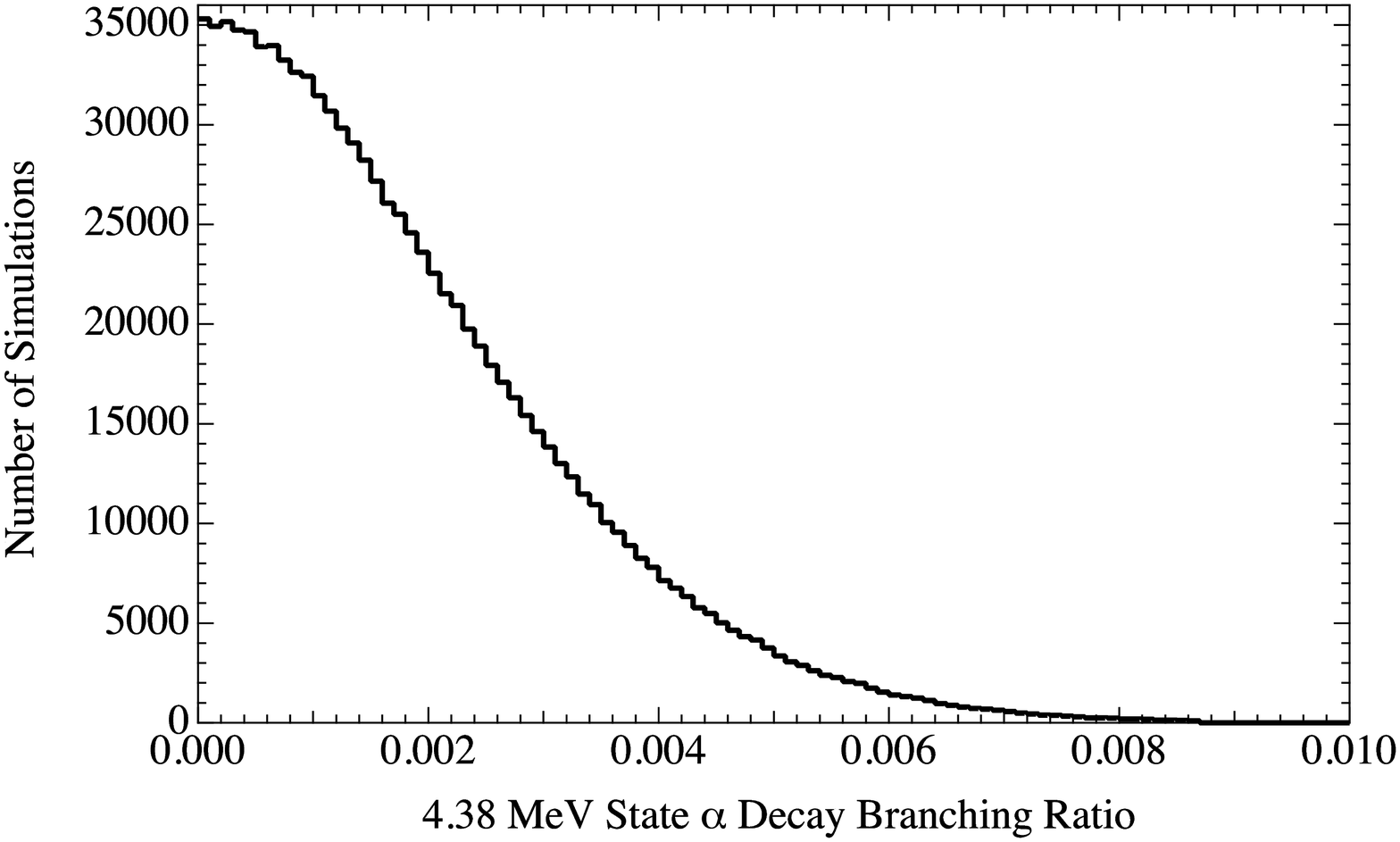}
\caption{Likelihood distribution of the $\alpha$ decay branching ratio of the 4.38 MeV state in $^{19}$Ne based on the measurement of Davids \emph{et al.} 2003.}
\label{ba438}
\end{figure}

We treat the 4.38 MeV state in the same way as we do the 4.03 MeV state, with the exception being that only two independent measurements of $\tau$ exist. The likelihood distribution of B$_\alpha$ for this state is taken from the measurement of \citet{davids03} and is shown in Figure \ref{ba438}.

The mean lifetimes of the 4.55 and 4.60 MeV states are drawn from the measurements of \citet{tan05} and \citet{mythili08}. The B$_\alpha$ values of the 4.55 and 4.60 MeV states are drawn from the measurements of \citet{magnus90}, \citet{laird02}, \citet{davids03}, and \citet{tan09}, with the weight of each measurement proportional to the inverse of its variance as usual.

In the cases of the 4.71 and 5.09 MeV states for which lifetime data are lacking we assume that isospin symmetry is valid and equate the reduced transition probabilities of these states with their analog states in $^{19}$F. Utilizing the lifetime measurements compiled in \cite{tilley95}, the $\gamma$ decay branching ratio measurement of \cite{pringle89}, and the resonance strength measurement of \cite{wilmes02}, we find that the radiative widths of the 4.71 MeV and 5.09 MeV states are $46^{+11}_{-8}$ meV and $110^{+110}_{-60}$ meV respectively. The B$_\alpha$ values are drawn from the measurements of \citet{magnus90}, \citet{davids03}, and \citet{tan09}.

\begin{figure}
\includegraphics[width=0.8\linewidth]{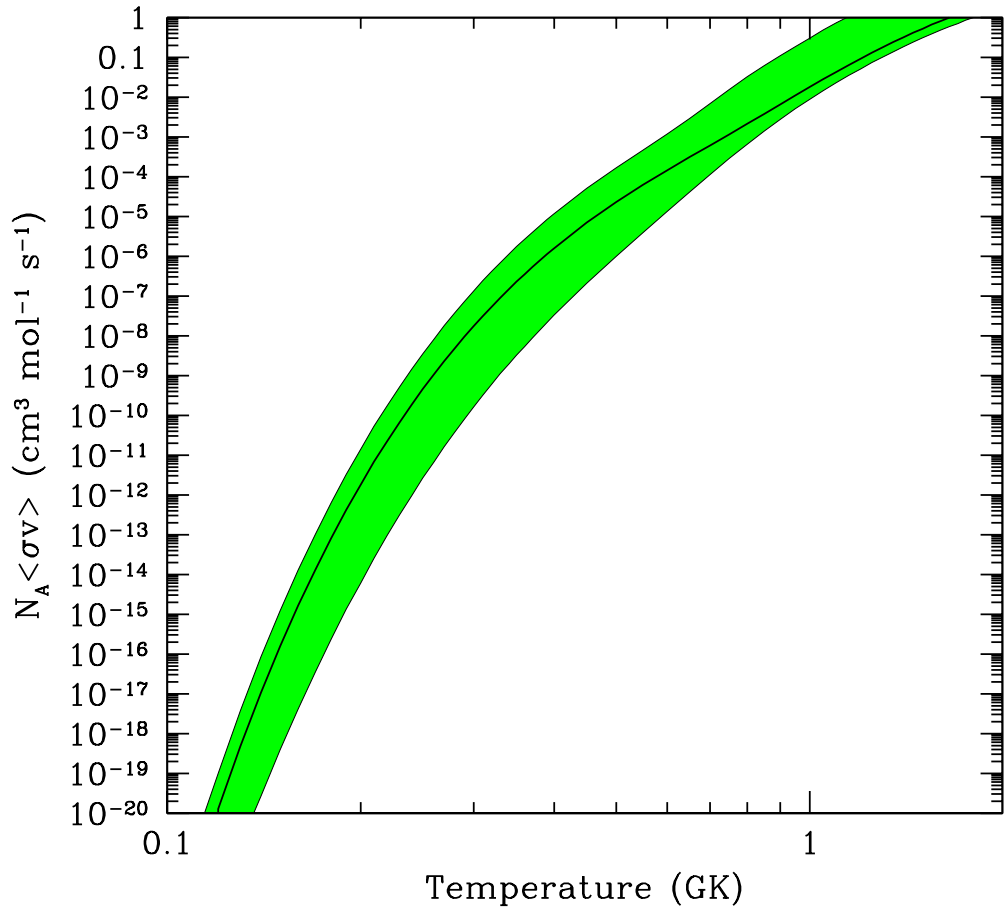}
\caption{Product of the Avogadro constant and the thermally averaged rate of the $^{15}$O($\alpha,\gamma)^{19}$Ne per particle pair as a function of temperature. Shown are the median and 99.73\% confidence level upper and lower limits of the probability distribution from the Monte Carlo simulation.}
\label{rateabs}
\end{figure}

Using the resonance parameter likelihood distributions described above, we calculated the thermally averaged rate of the $^{15}$O($\alpha,\gamma)^{19}$Ne per particle pair as a function of temperature one million times at a range of temperatures between 0.1 and 2 GK. Analysing the results, we determine upper and lower limits on the reaction rate at the 99.73\% confidence level and adopt the median of the calculated reaction rate distribution as our recommended value. The calculated reaction rates are shown in Figure \ref{rateabs} and numerical values are given in Table \ref{tabrates}.

\begin{figure}
\includegraphics[width=\linewidth]{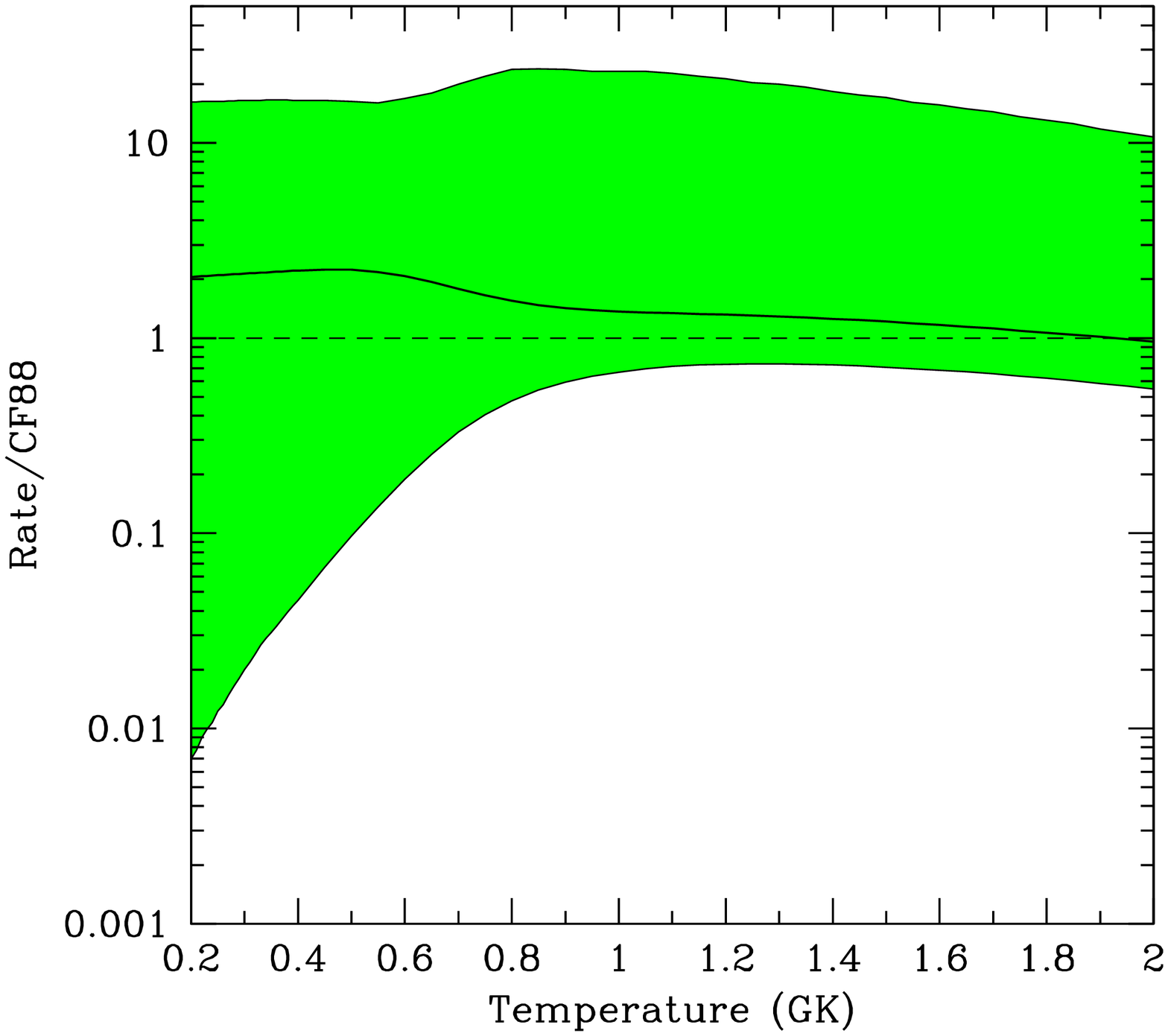}
\caption{Ratio of the Monte Carlo reaction rate to the reaction rate estimated in \citet{cf88}. The solid line is the ratio of the recommended reaction rate to that of \citet{cf88} and the shaded region is allowed at the 99.73\% confidence level.}
\label{ratecf88}
\end{figure}

\begin{figure}
\includegraphics[width=\linewidth]{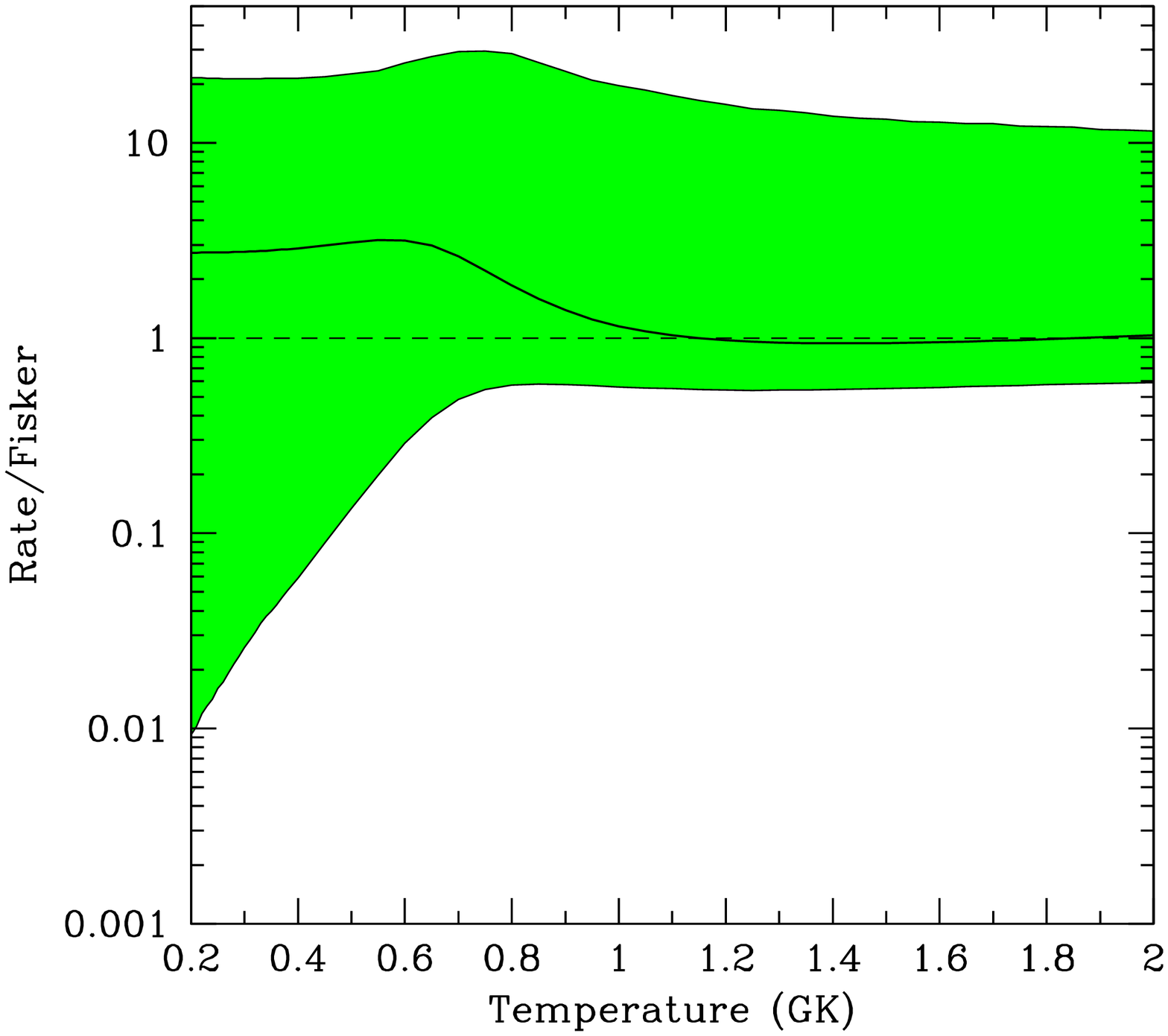}
\caption{Ratio of the Monte Carlo reaction rate to the reaction rate estimated in \citet{fisker06}. The solid line is the ratio of the recommended reaction rate to that of \citet{fisker06} and the shaded region is allowed at the 99.73\% confidence level.}
\label{ratefisker}
\end{figure}

Graphical comparisons of the Monte Carlo reaction rates to the reaction rates estimated in \citet{cf88} and \citet{fisker06} are shown in Figure \ref{ratecf88} and Figure \ref{ratefisker} respectively. In the temperature range from 0.1-2 GK the Monte Carlo lower limit on the reaction rate ranges between 2\% and 91\% of the lower limit of \citet{tan09}, the Monte Carlo central value lies between 0.71 and 1.04 times that of \citet{tan09}, and the Monte Carlo upper limit is between 3.2 and 9.4 times that of \citet{tan09}. Hence the central value and limits on the reaction rate estimated by \citet{tan09} fall well within the 99.73\% CL limits on the reaction rate obtained from the Monte Carlo simulation. Comparing to the recent reaction rate estimates of \citet{iliadis10}, we find that over the same temperature range of 0.1-2 GK our Monte Carlo lower limit on the reaction rate is between 0.03 and 3.6 times that of \citet{iliadis10}, our central value is between 1.7 and 3.3 times that of  \citet{iliadis10}, and our upper limit is between 7 and 14 times that of  \citet{iliadis10}. However, throughout the entire temperature range characterizing the initial phase of the thermonuclear runaway (below 0.7 GK), our Monte Carlo lower limit on the reaction rate is smaller than that of  \citet{iliadis10}.

\begin{table}
\caption{Product of the Avogadro constant and the thermally averaged rate of the $^{15}$O($\alpha,\gamma)^{19}$Ne reaction per particle pair in cm$^3$ mol$^{-1}$ s$^{-1}$\label{tabrates}}
\begin{tabular}{cccc}
\tableline\tableline
Temperature (GK) & 99.73\% CL Lower Limit & Median & 99.73\% CL Upper Limit\\
\tableline 
0.1 & $1.1\times10^{-25}$ & $1.0\times10^{-24}$ & $7.7\times10^{-24}$\\ 
0.2 & $6.1\times10^{-15}$ & $1.8\times10^{-12}$ & $1.4\times10^{-11}$\\ 
0.25 & $2.7\times10^{-12}$ & $4.6\times10^{-10}$ & $3.6\times10^{-9}$\\ 
0.3 & $1.6\times10^{-10}$ & $1.8\times10^{-8}$ & $1.4\times10^{-7}$\\
0.35 & $3.3\times10^{-9}$ & $2.3\times10^{-7}$ & $1.8\times10^{-6}$\\ 
0.4 & $3.2\times10^{-8}$ & $1.6\times10^{-6}$ & $1.2\times10^{-5}$\\ 
0.45 & $2.1\times10^{-7}$ & $7.0\times10^{-6}$ & $5.1\times10^{-5}$\\
0.5 & $1.0\times10^{-6}$ & $2.3\times10^{-5}$ & 0.00017   \\
0.6 & $1.3\times10^{-5}$ & 0.00015 & 0.0012\\ 
0.7 & 0.00011 & 0.00061 & 0.0069\\
0.8 & 0.00066 & 0.0021 & 0.033\\ 
0.9 & 0.0028 & 0.0066 & 0.11\\ 
1.0 & 0.0088 & 0.018 & 0.31\\ 
1.25 & 0.074 & 0.13 & 2.0\\ 
1.5 & 0.31 & 0.53 & 7.4\\
1.75 & 0.85 & 1.5 & 18\\ 
2.0 & 1.78 & 3.1 & 35\\ 
\tableline
\end{tabular}
\end{table}

\begin{figure}
\includegraphics[width=\linewidth]{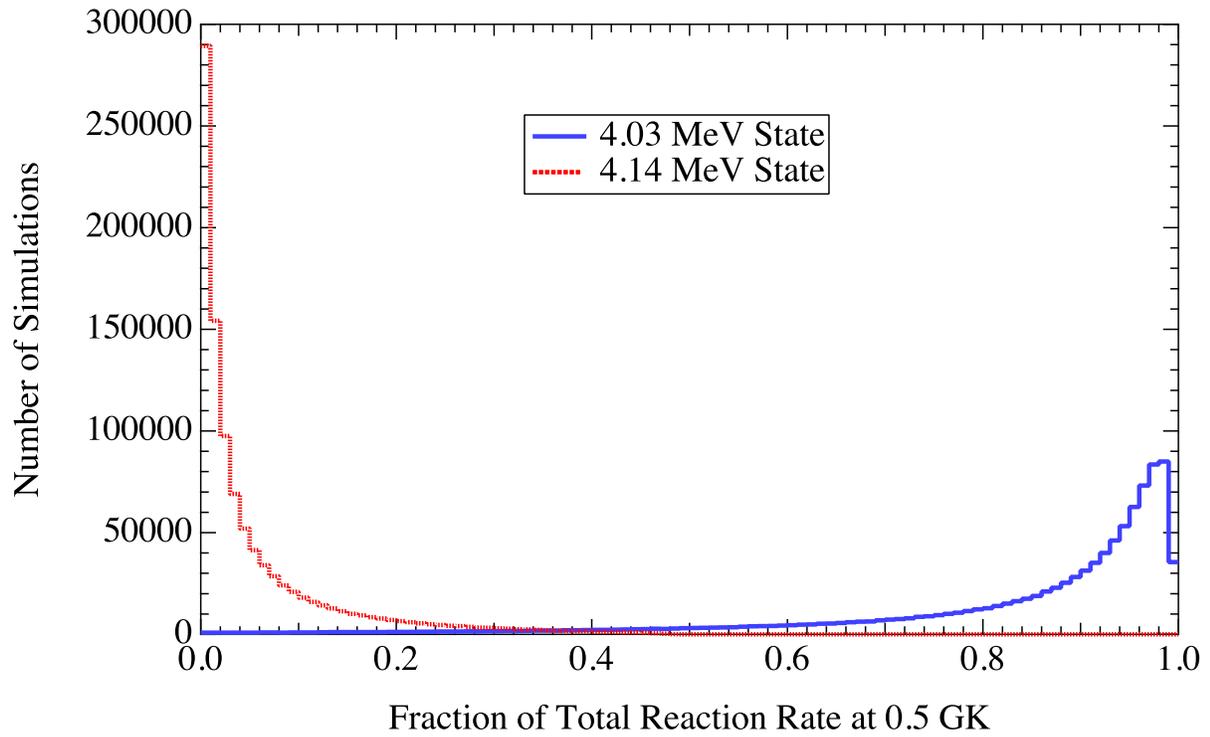}
\caption{Likelihood distributions of the fraction of the total reaction rate at 0.5 GK accounted for by the 4.03 and 4.14 MeV states. Below 0.6 GK, the reaction rate is likely dominated by the contribution of the 4.03 MeV state with small contributions from the 4.14 and 4.20 MeV states.}
\label{frac403414}
\end{figure}

Given the large uncertainties in the B$_\alpha$ values of the low lying states, it is impossible to precisely specify the fraction of the total reaction rate accounted for by the resonant contribution of any particular state. But in the Monte Carlo approach taken here it is possible to find the likelihood of a given state accounting for any definite fraction of the total reaction rate. Figure \ref{frac403414} shows the likelihood distributions of the fraction of the total reaction rate at 0.5 GK accounted for by the resonant contributions of the 4.03 and 4.14 MeV states. It is possible to say that at this temperature and indeed below 0.6 GK, the 4.03 MeV state likely dominates the reaction rate.

\section{Simulations of Type I X-Ray Bursts}

In order to assess the impact of the current uncertainties in the $^{15}$O($\alpha$, $\gamma$) rate on Type I x-ray burst (XRB) models, we have performed a new set of simulations with SHIVA, a one dimensional (spherically symmetric), hydrodynamic, implicit, Lagrangian code used extensively in the modeling of stellar explosions such as classical novae \citep{josehernanz98} and XRB's 
\citep{jose10}. To this end, three different models of thermonuclear bursts, with identical input physics except for the specific choice of
the $^{15}$O($\alpha$, $\gamma$)$^{19}$Ne rate have been computed: the lower limit reaction rate dubbed Model A, the recommended rate Model B, and the upper limit Model C. In all cases the accretion of solar-like matter with hydrogen mass fraction X = 0.7048, helium mass fraction Y = 0.2752, and metallicity Z = 0.02 onto a 1.4 $M_\odot$ neutron star (with initial luminosity $L_{i} = 1.6 \times 10^{34}$ erg s$^{-1}$ = 4.14 $L_\odot$), at a mass accretion rate $\dot M_{acc} = 1.75 \times 10^{-9}$ $M_\odot$ yr$^{-1}$ 
(corresponding to 0.08 $\dot M_{Edd}$) has been adopted. The Eddington accretion rate $\dot M_{Edd}$ is the accretion rate that produces a luminosity at which the force due to radiation pressure equals the local gravitational force. All metals are initially assumed to be in the form of $^{14}$N, which is justified considering the rapidity with which the CNO nuclei are converted to $^{14}$N at the high temperatures reached early in the burst and the negligible energy this releases compared with accretion \citep{woosley04}. The model adopted in this work is qualitatively similar to model zM computed by \citet{woosley04} in the framework of the 1D, hydrodynamic, implicit code KEPLER, as well as to the model computed by \citet{fisker06}. We note however that whereas \citet{woosley04} assume a value of 10 km for the (Newtonian) neutron star radius, our model yields a value of 13.1~km following the integration of the neutron star structure. In turn, the calculations reported by \citet{fisker06}, in a general relativistic framework, relied on a radius of 11 km for the same neutron star mass. Differences in the neutron star size and thereby surface gravity may affect the strength of the explosion through the accreted mass, peak temperature, nucleosynthesis, etc. to some extent.

The code has been linked to a fully updated nuclear reaction network containing 324 nuclides from $^1$H to $^{107}$Te and 1392 nuclear processes, and includes the most important charged particle induced reactions occurring among the nuclides between $^1$H and $^{107}$Te as well as their corresponding inverse processes (see \citet{parikh08} and \citet{jose10} for details).
Neutron captures are not taken into account since they play a very minor role in XRB nucleosynthesis. SHIVA uses a time-dependent formalism for convective transport whenever the characteristic convective timescale becomes larger than the integration time step. Partial mixing between adjacent convective shells is treated by means of a diffusion equation \citep{prialnik79}. Accretion is computed by redistributing material through a constant number of envelope shells \citep{kuttersparks80}: a thin envelope, containing $1.1 \times 10^{18}$ g of material (less than 1/1000 of the total envelope mass accreted during the first bursting episode), equally distributed through all the envelope shells, is chosen as an initial condition. The model is then relaxed using a few very large time steps to guarantee hydrostatic equilibrium. More details on the adopted input physics can be found in \citet{jose10}.

\section{Results and Discussion}

\begin{figure}
\includegraphics[width=\linewidth]{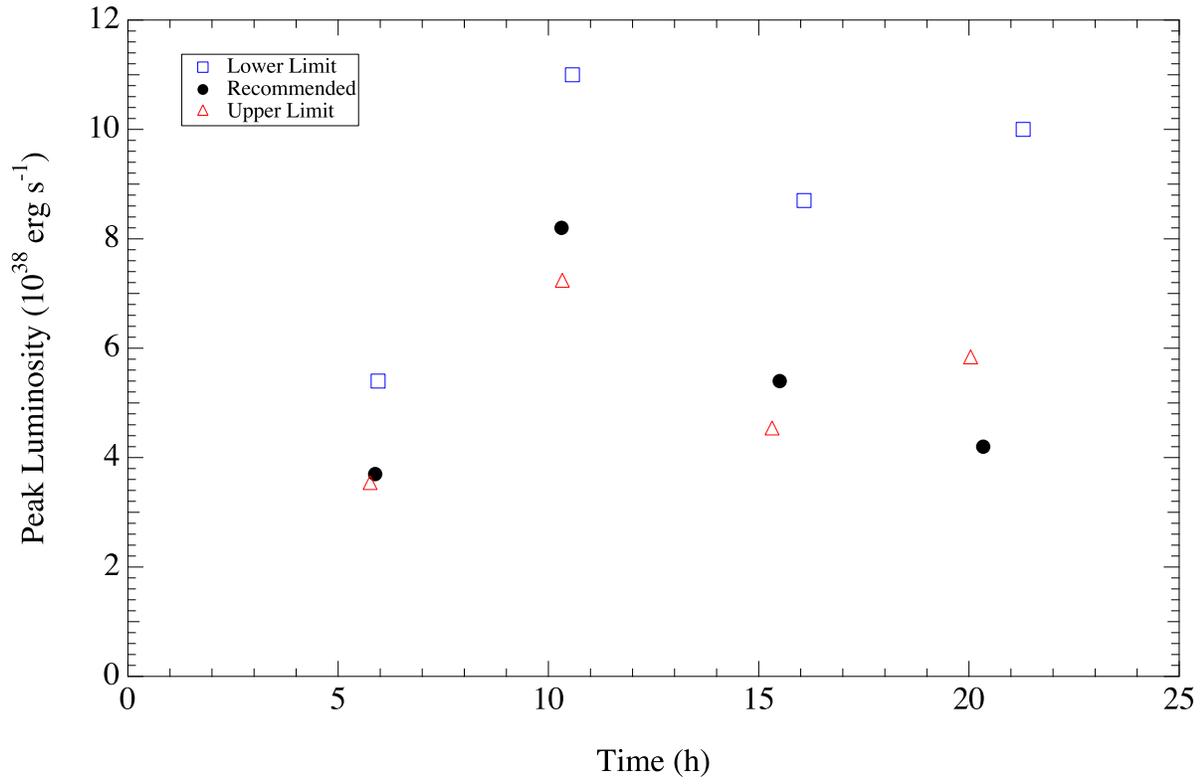}
\caption{Peak luminosities attained in the first four x-ray bursts and the time they were reached in the model calculations using three different $^{15}$O($\alpha$,$\gamma)^{19}$Ne reaction rates.}
\label{bursts}
\end{figure}

A sequence of four consecutive bursts has been computed for all three models reported in this work. Summaries of the gross properties of the different sequences are given in Figure \ref{bursts}, Table \ref{tablea}, Table \ref{tableb} and Table \ref{tablec}. For each model, the mean mass fractions of all nuclides $i$ that are stable or have a half-life $>$1 h and are present with a mass fraction $X_i > 10^{-9}$ in the envelope at the end of the  fourth burst are given in Table \ref{tablenuc}.

As a framework for our discussion, we present first a summary of the results for Model B, in which the recommended $^{15}$O($\alpha$,$\gamma$)$^{19}$Ne rate has been adopted. As shown in Table \ref{tableb}, peak temperatures $T_{peak} \sim 1.1-1.3$ GK and peak luminosities $L_{peak} \sim (1-2) \times 10^5$ $L_\odot$ are reached in this model. Recurrence times between bursts around $\tau_{rec} \sim$ 4.5-6 hr and $\alpha \sim 28-50$, have also been obtained. The ratio between persistent and burst luminosities, $\alpha$, is defined as \begin{equation}\alpha = \frac{\int_t^{t+\tau_{rec}} L(t)\,dt}{\int_{t'}^{t'+\tau_{0.01}} L(t)\,dt},\end{equation} with the latter integrated over the time during which the burst exceeds 1\% of its peak luminosity, $\tau_{0.01}$. During the interburst period, the accretion luminosity, $L_{acc} = G M \dot{M} / R \sim 1.5 \times 10^{37}$ erg s$^{-1}$, will hide the thermal emission from the cooling ashes.

The values reported for Model B are qualitatively in agreement with those inferred from several XRB sources,
such as the {\it textbook burster} GS 1826-24 [$\tau_{rec} = 5.74 \pm 0.13$ h, $\alpha = 41.7 \pm 1.6$], 4U 1323-62 [$\tau_{rec} = 5.3$ h, $\alpha = 38 \pm 4$], or 4U 1608-52 [$\tau_{rec} =$ 4.14-7.5 h, $\alpha =$ 41-54], to quote a few examples from \citet{galloway08}. From the nucleosynthesis viewpoint,  the main nuclear flow is dominated by the rp-process (rapid proton-captures and $\beta^+$-decays), the 3$\alpha$-reaction, and the $\alpha$p-process (a sequence of ($\alpha$, p) and (p, $\gamma$) reactions). The main flow proceeds away from the valley of stability, eventually reaching the proton drip line beyond A = 38. In this context, much of the initial H and $^4$He is transformed into heavier nuclei. However, it is important to stress the presence of  some unburned H (X = 0.05) and $^4$He (Y = 0.11) in the envelope at the end of the fourth burst (see Table \ref{tablenuc}). Indeed, the mean, mass-averaged metallicity at the end of the fourth bursting episode has increased from the original (accreted) metallicity of $Z_{i} \sim 0.02$ to a value of $Z \sim 0.84$. The nuclei in the envelope with the largest mass fractions are $^{60}$Ni (0.29), $^{64}$Zn (0.14), $^{4}$He (0.11), $^{32}$S (0.09), and $^{56}$Ni (0.06). A mass-averaged $^{12}$C yield of $\sim 0.02$ is obtained at the end of the fourth burst, not enough to power a superburst (which requires X($^{12}$C)$_{min} \sim 0.1$ at the envelope base; see \citet{cumming01,strohmayer02,brown04,cooper05,cumming05,cooper06}). The corresponding nucleosynthetic endpoint (defined by the heaviest isotope with a mass fraction $X_i > 10^{-9}$) is found around $^{100}$Pd.

\begin{deluxetable}{ccccc}
\small
\tablecaption{ Summary of burst properties for Model A. }
                  \tablehead{
		  \colhead{Burst number}   &
		  \colhead{T$_{peak}$ (GK)}   &
		  \colhead{$\tau_{rec}$ (h)} &
		  \colhead{L$_{peak}$ (L$_\odot$)} &
		  \colhead{$\alpha$} 
		                    }
\startdata
 1    & $1.10$& 5.95        & $1.4 \times 10^5$&  50 \\
 2    & $1.25$& 4.63        & $2.9 \times 10^5$&  29 \\
 3    & $1.21$& 5.51        & $2.3 \times 10^5$&  38 \\
 4    & $1.21$& 5.21        & $2.7 \times 10^5$&  30 \\ [0.5ex]
\enddata
\label{tablea}
\end{deluxetable}

\begin{deluxetable}{ccccc}
\small
\tablecaption{ Summary of burst properties for Model B. }
                  \tablehead{
		  \colhead{Burst number}   &
		  \colhead{T$_{peak}$ (GK)}   &
		  \colhead{$\tau_{rec}$ (h)} &
		  \colhead{L$_{peak}$ (L$_\odot$)} &
		  \colhead{$\alpha$} 
		                    }
\startdata
 1    & $1.06$& 5.88        & $9.7 \times 10^4$&  50 \\
 2    & $1.25$& 4.44        & $2.1 \times 10^5$&  28 \\
 3    & $1.14$& 5.19        & $1.4 \times 10^5$&  34 \\
 4    & $1.12$& 4.84        & $1.1 \times 10^5$&  32 \\ [0.5ex]
\enddata
\label{tableb}
\end{deluxetable}

\begin{deluxetable}{ccccc}
\small
\tablecaption{ Summary of burst properties for Model C. }
                  \tablehead{
		  \colhead{Burst number}   &
		  \colhead{T$_{peak}$ (GK)}   &
		  \colhead{$\tau_{rec}$ (h)} &
		  \colhead{L$_{peak}$ (L$_\odot$)} &
		  \colhead{$\alpha$} 
		                    }
\startdata
 1    & $1.03$& 5.77        & $7.7 \times 10^4$&  52 \\
 2    & $1.26$& 4.57        & $1.9 \times 10^5$&  27 \\
 3    & $1.17$& 5.00        & $1.2 \times 10^5$&  33 \\
 4    & $1.17$& 4.72        & $1.5 \times 10^5$&  29 \\ [0.5ex]
\enddata
\label{tablec}
\end{deluxetable}

\begin{deluxetable}{cccc}
\small
\tablecaption{Mean mass fractions of the envelope at the end of the fourth burst, for stable nuclides and those with half lives over 1 h and mass fractions X$_i > 10^{-9}$.}
                 \tablehead{
		 \colhead{Nucleus}   &
		 \colhead{Model A}   &
		 \colhead{Model B}   &
		 \colhead{Model C}
		          }
\startdata
   $^1$H     & $3.2 \times 10^{-2}$   &  $4.9 \times 10^{-2}$      &  $4.4 \times 10^{-2}$        \\
   $^4$He    & $9.7 \times 10^{-2}$   &  $1.1 \times 10^{-1}$      &  $1.0 \times 10^{-1}$        \\ 
   $^{12}$C  & $1.5 \times 10^{-2}$   &  $1.9 \times 10^{-2}$      &  $1.7 \times 10^{-2}$        \\
   $^{13}$C  & $1.3 \times 10^{-4}$   &  $1.5 \times 10^{-4}$      &  $1.3 \times 10^{-4}$        \\
   $^{14}$N  & $7.8 \times 10^{-4}$   &  $8.4 \times 10^{-4}$      &  $7.6 \times 10^{-4}$        \\
   $^{15}$N  & $9.3 \times 10^{-4}$   &  $1.0 \times 10^{-3}$      &  $8.2 \times 10^{-4}$        \\
   $^{16}$O  & $2.1 \times 10^{-4}$   &  $2.6 \times 10^{-4}$      &  $2.5 \times 10^{-4}$        \\
   $^{17}$O  & $1.1 \times 10^{-6}$   &  $8.2 \times 10^{-6}$      &  $6.6 \times 10^{-6}$        \\
   $^{18}$O  & $1.7 \times 10^{-5}$   &  $7.0 \times 10^{-6}$      &  $9.8 \times 10^{-6}$        \\
   $^{18}$F  & $3.6 \times 10^{-5}$   &  $1.9 \times 10^{-5}$      &  $2.5 \times 10^{-5}$        \\
   $^{19}$F  & $4.7 \times 10^{-5}$   &  $4.5 \times 10^{-5}$      &  $3.5 \times 10^{-5}$        \\
   $^{20}$Ne & $1.6 \times 10^{-4}$   &  $4.2 \times 10^{-4}$      &  $3.7 \times 10^{-4}$        \\
   $^{21}$Ne & $2.5 \times 10^{-6}$   &  $4.3 \times 10^{-6}$      &  $4.8 \times 10^{-6}$        \\
   $^{22}$Ne & $6.6 \times 10^{-6}$   &  $2.5 \times 10^{-5}$      &  $1.8 \times 10^{-5}$        \\
   $^{22}$Na & $6.0 \times 10^{-4}$   &  $6.8 \times 10^{-4}$      &  $9.0 \times 10^{-4}$        \\
   $^{23}$Na & $3.8 \times 10^{-5}$   &  $1.9 \times 10^{-4}$      &  $1.2 \times 10^{-4}$        \\
   $^{24}$Mg & $4.1 \times 10^{-4}$   &  $1.4 \times 10^{-3}$      &  $8.9 \times 10^{-4}$        \\
   $^{25}$Mg & $5.0 \times 10^{-4}$   &  $1.6 \times 10^{-3}$      &  $1.1 \times 10^{-3}$        \\
   $^{26}$Mg & $1.1 \times 10^{-3}$   &  $1.0 \times 10^{-3}$      &  $1.1 \times 10^{-3}$        \\
$^{26}$Al$^g$& $1.3 \times 10^{-5}$   &  $1.3 \times 10^{-4}$      &  $6.6 \times 10^{-5}$        \\
   $^{27}$Al & $3.9 \times 10^{-4}$   &  $7.0 \times 10^{-4}$      &  $7.8 \times 10^{-4}$        \\
   $^{28}$Si & $5.5 \times 10^{-3}$   &  $1.6 \times 10^{-2}$      &  $8.8 \times 10^{-3}$        \\
   $^{29}$Si & $4.5 \times 10^{-4}$   &  $7.7 \times 10^{-4}$      &  $5.3 \times 10^{-4}$        \\
   $^{30}$Si & $2.3 \times 10^{-3}$   &  $3.0 \times 10^{-3}$      &  $3.0 \times 10^{-3}$        \\
   $^{31}$P  & $1.2 \times 10^{-3}$   &  $1.7 \times 10^{-3}$      &  $1.4 \times 10^{-3}$        \\
   $^{32}$S  & $8.7 \times 10^{-2}$   &  $9.0 \times 10^{-2}$      &  $7.3 \times 10^{-2}$        \\
   $^{33}$S  & $4.3 \times 10^{-3}$   &  $4.6 \times 10^{-3}$      &  $4.0 \times 10^{-3}$        \\
   $^{34}$S  & $1.3 \times 10^{-2}$   &  $1.5 \times 10^{-2}$      &  $1.3 \times 10^{-2}$        \\
   $^{35}$Cl & $1.4 \times 10^{-2}$   &  $6.3 \times 10^{-3}$      &  $7.3 \times 10^{-3}$        \\
   $^{36}$Ar & $1.6 \times 10^{-2}$   &  $7.0 \times 10^{-3}$      &  $7.7 \times 10^{-3}$        \\
   $^{37}$Cl & $3.2 \times 10^{-6}$   &  $2.5 \times 10^{-6}$      &  $1.6 \times 10^{-6}$        \\ 
   $^{37}$Ar & $1.2 \times 10^{-3}$   &  $7.4 \times 10^{-4}$      &  $7.6 \times 10^{-4}$        \\
   $^{38}$Ar & $9.5 \times 10^{-3}$   &  $6.2 \times 10^{-3}$      &  $6.8 \times 10^{-3}$        \\
   $^{39}$K  & $1.5 \times 10^{-2}$   &  $1.1 \times 10^{-2}$      &  $1.1 \times 10^{-2}$        \\
   $^{40}$Ca & $5.7 \times 10^{-3}$   &  $4.6 \times 10^{-3}$      &  $4.0 \times 10^{-3}$        \\
   $^{41}$K  & $1.2 \times 10^{-7}$   &  $3.5 \times 10^{-8}$      &  $3.5 \times 10^{-8}$        \\
   $^{41}$Ca & $6.0 \times 10^{-5}$   &  $7.3 \times 10^{-5}$      &  $6.3 \times 10^{-5}$        \\
   $^{42}$Ca & $2.2 \times 10^{-3}$   &  $1.5 \times 10^{-3}$      &  $1.4 \times 10^{-3}$        \\
   $^{43}$Ca & $1.4 \times 10^{-3}$   &  $1.3 \times 10^{-3}$      &  $1.2 \times 10^{-3}$        \\
   $^{43}$Sc & $1.3 \times 10^{-3}$   &  $8.2 \times 10^{-4}$      &  $9.5 \times 10^{-4}$        \\
   $^{44}$Ca & $7.8 \times 10^{-5}$   &  $3.1 \times 10^{-5}$      &  $2.2 \times 10^{-5}$        \\
   $^{44}$Sc & $8.8 \times 10^{-5}$   &  $2.7 \times 10^{-5}$      &  $4.5 \times 10^{-5}$        \\
   $^{44}$Ti & $7.4 \times 10^{-4}$   &  $6.8 \times 10^{-4}$      &  $5.7 \times 10^{-4}$        \\
   $^{45}$Sc & $9.6 \times 10^{-5}$   &  $1.3 \times 10^{-4}$      &  $8.9 \times 10^{-5}$        \\
   $^{45}$Ti & $7.5 \times 10^{-5}$   &  $9.7 \times 10^{-5}$      &  $9.7 \times 10^{-5}$        \\
   $^{46}$Ti & $3.3 \times 10^{-3}$   &  $2.5 \times 10^{-3}$      &  $2.1 \times 10^{-3}$        \\
   $^{47}$Ti & $9.8 \times 10^{-4}$   &  $8.6 \times 10^{-4}$      &  $7.1 \times 10^{-4}$        \\
   $^{48}$Ti & $3.5 \times 10^{-6}$   &  $3.2 \times 10^{-6}$      &  $2.1 \times 10^{-6}$        \\
   $^{48}$V  & $3.8 \times 10^{-4}$   &  $3.9 \times 10^{-4}$      &  $2.6 \times 10^{-4}$        \\
   $^{48}$Cr & $1.9 \times 10^{-3}$   &  $2.0 \times 10^{-3}$      &  $1.7 \times 10^{-3}$        \\
   $^{49}$Ti & $6.7 \times 10^{-7}$   &  $7.9 \times 10^{-7}$      &  $7.2 \times 10^{-7}$        \\
   $^{49}$V  & $1.1 \times 10^{-3}$   &  $1.4 \times 10^{-3}$      &  $1.2 \times 10^{-3}$        \\
   $^{50}$Cr & $3.4 \times 10^{-3}$   &  $2.9 \times 10^{-3}$      &  $2.4 \times 10^{-3}$        \\
   $^{51}$V  & $3.4 \times 10^{-5}$   &  $3.5 \times 10^{-5}$      &  $2.7 \times 10^{-5}$        \\
   $^{51}$Cr & $6.7 \times 10^{-3}$   &  $6.1 \times 10^{-3}$      &  $5.2 \times 10^{-3}$        \\
   $^{52}$Cr & $6.9 \times 10^{-5}$   &  $9.2 \times 10^{-5}$      &  $6.5 \times 10^{-5}$        \\
   $^{52}$Mn & $2.9 \times 10^{-3}$   &  $4.8 \times 10^{-3}$      &  $3.2 \times 10^{-3}$        \\
   $^{52}$Fe & $6.9 \times 10^{-3}$   &  $1.2 \times 10^{-2}$      &  $9.3 \times 10^{-3}$        \\
   $^{53}$Mn & $7.7 \times 10^{-4}$   &  $1.1 \times 10^{-3}$      &  $7.9 \times 10^{-4}$        \\
   $^{54}$Fe & $2.0 \times 10^{-3}$   &  $1.7 \times 10^{-3}$      &  $1.3 \times 10^{-3}$        \\
   $^{55}$Mn & $1.8 \times 10^{-7}$   &  $1.5 \times 10^{-7}$      &  $1.2 \times 10^{-7}$        \\
   $^{55}$Fe & $1.2 \times 10^{-3}$   &  $1.1 \times 10^{-3}$      &  $8.4 \times 10^{-4}$        \\
   $^{55}$Co & $4.7 \times 10^{-3}$   &  $4.1 \times 10^{-3}$      &  $3.4 \times 10^{-3}$        \\
   $^{56}$Fe & $7.2 \times 10^{-6}$   &  $3.6 \times 10^{-6}$      &  $3.2 \times 10^{-6}$        \\
   $^{56}$Co & $3.8 \times 10^{-3}$   &  $2.1 \times 10^{-3}$      &  $1.9 \times 10^{-3}$        \\
   $^{56}$Ni & $1.4 \times 10^{-1}$   &  $6.3 \times 10^{-2}$      &  $6.0 \times 10^{-2}$        \\
   $^{57}$Fe & $4.7 \times 10^{-7}$   &  $1.9 \times 10^{-7}$      &  $1.6 \times 10^{-7}$        \\
   $^{57}$Co & $4.7 \times 10^{-4}$   &  $2.6 \times 10^{-4}$      &  $1.5 \times 10^{-4}$        \\
   $^{57}$Ni & $8.5 \times 10^{-3}$   &  $3.3 \times 10^{-3}$      &  $2.1 \times 10^{-3}$        \\
   $^{58}$Ni & $4.2 \times 10^{-3}$   &  $2.5 \times 10^{-3}$      &  $1.6 \times 10^{-3}$        \\
   $^{59}$Ni & $5.5 \times 10^{-3}$   &  $4.1 \times 10^{-3}$      &  $2.7 \times 10^{-3}$        \\
   $^{60}$Ni & $2.6 \times 10^{-1}$   &  $2.9 \times 10^{-1}$      &  $2.9 \times 10^{-1}$        \\
   $^{61}$Ni & $1.6 \times 10^{-3}$   &  $2.7 \times 10^{-3}$      &  $1.5 \times 10^{-3}$        \\
   $^{61}$Cu & $1.1 \times 10^{-3}$   &  $3.8 \times 10^{-3}$      &  $2.8 \times 10^{-3}$        \\
   $^{62}$Ni & $4.3 \times 10^{-4}$   &  $5.3 \times 10^{-4}$      &  $3.1 \times 10^{-4}$        \\
   $^{62}$Zn & $7.9 \times 10^{-4}$   &  $1.7 \times 10^{-3}$      &  $1.1 \times 10^{-3}$        \\
   $^{63}$Cu & $2.4 \times 10^{-3}$   &  $3.1 \times 10^{-3}$      &  $2.3 \times 10^{-3}$        \\
   $^{64}$Zn & $1.4 \times 10^{-1}$   &  $1.4 \times 10^{-1}$      &  $1.7 \times 10^{-1}$        \\
   $^{65}$Cu & $7.5 \times 10^{-7}$   &  $1.5 \times 10^{-6}$      &  $1.2 \times 10^{-6}$        \\
   $^{65}$Zn & $1.2 \times 10^{-3}$   &  $2.8 \times 10^{-3}$      &  $2.4 \times 10^{-3}$        \\
   $^{66}$Zn & $2.2 \times 10^{-4}$   &  $4.0 \times 10^{-4}$      &  $3.3 \times 10^{-4}$        \\
   $^{66}$Ga & $4.7 \times 10^{-4}$   &  $1.0 \times 10^{-3}$      &  $8.1 \times 10^{-4}$        \\
   $^{66}$Ge & $3.3 \times 10^{-4}$   &  $8.7 \times 10^{-4}$      &  $8.4 \times 10^{-4}$        \\
   $^{67}$Ga & $8.0 \times 10^{-4}$   &  $1.4 \times 10^{-3}$      &  $1.3 \times 10^{-3}$        \\
   $^{68}$Ge & $4.4 \times 10^{-2}$   &  $4.7 \times 10^{-2}$      &  $5.9 \times 10^{-2}$        \\
   $^{69}$Ge & $9.5 \times 10^{-4}$   &  $2.3 \times 10^{-3}$      &  $2.3 \times 10^{-3}$        \\
   $^{70}$Ge & $2.9 \times 10^{-4}$   &  $8.1 \times 10^{-4}$      &  $7.5 \times 10^{-4}$        \\
   $^{71}$As & $4.5 \times 10^{-4}$   &  $9.9 \times 10^{-4}$      &  $1.1 \times 10^{-3}$        \\
   $^{72}$Se & $1.4 \times 10^{-2}$   &  $1.6 \times 10^{-2}$      &  $2.2 \times 10^{-2}$        \\
   $^{73}$Se & $4.8 \times 10^{-4}$   &  $1.3 \times 10^{-3}$      &  $1.4 \times 10^{-3}$        \\
   $^{74}$Se & $2.7 \times 10^{-4}$   &  $6.3 \times 10^{-4}$      &  $7.3 \times 10^{-4}$        \\
   $^{75}$Br & $2.3 \times 10^{-4}$   &  $6.7 \times 10^{-4}$      &  $7.7 \times 10^{-4}$        \\
   $^{76}$Kr & $4.9 \times 10^{-3}$   &  $6.1 \times 10^{-3}$      &  $9.2 \times 10^{-3}$        \\
   $^{77}$Kr & $3.0 \times 10^{-4}$   &  $7.8 \times 10^{-4}$      &  $1.0 \times 10^{-3}$        \\
   $^{78}$Kr & $1.9 \times 10^{-4}$   &  $5.4 \times 10^{-4}$      &  $7.2 \times 10^{-4}$        \\
   $^{79}$Kr & $1.3 \times 10^{-4}$   &  $3.5 \times 10^{-4}$      &  $4.8 \times 10^{-4}$        \\
   $^{80}$Sr & $1.8 \times 10^{-3}$   &  $2.5 \times 10^{-3}$      &  $4.1 \times 10^{-3}$        \\
   $^{81}$Rb & $2.2 \times 10^{-4}$   &  $5.5 \times 10^{-4}$      &  $8.5 \times 10^{-4}$        \\
   $^{82}$Sr & $2.6 \times 10^{-4}$   &  $7.5 \times 10^{-4}$      &  $1.1 \times 10^{-3}$        \\
   $^{83}$Sr & $1.7 \times 10^{-4}$   &  $4.3 \times 10^{-4}$      &  $7.6 \times 10^{-4}$        \\
   $^{84}$Sr & $6.8 \times 10^{-5}$   &  $3.8 \times 10^{-4}$      &  $4.6 \times 10^{-4}$        \\
   $^{85}$Y  & $1.9 \times 10^{-4}$   &  $4.6 \times 10^{-4}$      &  $6.9 \times 10^{-4}$        \\
   $^{86}$Zr & $2.1 \times 10^{-4}$   &  $6.3 \times 10^{-4}$      &  $9.8 \times 10^{-4}$        \\
   $^{87}$Zr & $1.6 \times 10^{-4}$   &  $5.4 \times 10^{-4}$      &  $9.4 \times 10^{-4}$        \\
   $^{88}$Zr & $2.5 \times 10^{-4}$   &  $2.8 \times 10^{-4}$      &  $7.3 \times 10^{-4}$        \\
   $^{89}$Nb & $6.5 \times 10^{-4}$   &  $6.1 \times 10^{-4}$      &  $2.4 \times 10^{-3}$        \\
   $^{90}$Mo & $1.2 \times 10^{-4}$   &  $2.7 \times 10^{-4}$      &  $1.0 \times 10^{-3}$        \\
   $^{91}$Nb & $3.7 \times 10^{-5}$   &  $1.5 \times 10^{-4}$      &  $5.8 \times 10^{-4}$        \\
   $^{92}$Mo & $9.2 \times 10^{-6}$   &  $7.1 \times 10^{-5}$      &  $2.6 \times 10^{-4}$        \\
   $^{93}$Tc & $1.4 \times 10^{-5}$   &  $7.7 \times 10^{-5}$      &  $2.6 \times 10^{-4}$        \\
   $^{94}$Tc & $1.7 \times 10^{-5}$   &  $6.9 \times 10^{-5}$      &  $3.0 \times 10^{-4}$        \\
   $^{95}$Ru & $7.7 \times 10^{-7}$   &  $8.4 \times 10^{-6}$      &  $5.3 \times 10^{-5}$        \\
   $^{96}$Ru & $1.7 \times 10^{-7}$   &  $1.5 \times 10^{-6}$      &  $7.6 \times 10^{-6}$        \\
   $^{97}$Ru & $1.7 \times 10^{-7}$   &  $6.6 \times 10^{-7}$      &  $5.8 \times 10^{-6}$        \\
   $^{98}$Ru & $5.2 \times 10^{-8}$   &  $1.5 \times 10^{-7}$      &  $2.6 \times 10^{-6}$        \\
   $^{99}$Rh & $1.3 \times 10^{-8}$   &  $2.2 \times 10^{-8}$      &  $8.5 \times 10^{-7}$        \\
  $^{100}$Pd & $2.4 \times 10^{-9}$   &  $3.3 \times 10^{-9}$      &  $2.1 \times 10^{-7}$        \\
  $^{101}$Pd & $1.4 \times 10^{-9}$   &  -                         &  $9.7 \times 10^{-8}$        \\
  $^{102}$Pd & -                      &  -                         &  $1.6 \times 10^{-8}$        \\
  $^{103}$Ag & -                      &  -                         &  $1.5 \times 10^{-9}$        \\[0.5ex]
\enddata
\label{tablenuc}
\end{deluxetable}

When the lower limit for the $^{15}$O($\alpha$,$\gamma$)$^{19}$Ne reaction rate is adopted (Model A), peak temperatures and luminosities of $T_{peak} \sim 1.1-1.3$ GK and $L_{peak} \sim (1-3) \times 10^5$  $L_\odot$ are found (see Table \ref{tablea}). Recurrence times between bursts around $\tau_{rec} \sim$ 4.5-6 h and ratios between persistent and burst luminosities, $\alpha \sim 29-50$, have also been obtained. All in all, these values are very similar to those found with Model B. But more important, the model exhibits quasi-periodic bursting episodes in sharp contrast with the results reported in a similar analysis by \citet{fisker06}. In that work, the adopted lower limit on the reaction rate, which happens to lie within a factor of two from the lower limit found in the present Monte Carlo simulation, led to steady state burning of the accreted layer without bursting. With respect to nucleosynthesis, similar amounts of unburned H (0.03, by mass) and $^4$He (0.1) are found at the end of the fourth burst (see Table \ref{tablenuc}). The mean, mass-averaged metallicity at the end of the fourth bursting episode reached a value of $Z \sim 0.87$ in this model. The most abundant nuclides in the envelope in this model are $^{60}$Ni (0.26), $^{64}$Zn (0.14), $^{56}$Ni (0.14), $^{4}$He (0.1), and $^{32}$S (0.09). A mass-averaged $^{12}$C yield of $\sim 0.02$ is also obtained at the end of the fourth burst, while the nucleosynthetic endpoint corresponds to $^{101}$Pd.

Finally, when the upper limit for the $^{15}$O($\alpha$,$\gamma$) rate is adopted (Model C), peak temperatures and luminosities of $T_{peak} \sim 1-1.3$ GK and $L_{peak} \sim (0.8-2) \times 10^5$ $L_\odot$, are found (with the lower value obtained for the first burst, interpreted as driven by the initial conditions; see Table \ref{tablec}). Recurrence times between bursts around $\tau_{rec} \sim$ 4.5-6 h and ratios between persistent and burst luminosities, $\alpha \sim 27-52$, have also been obtained. In summary, the results are very similar to those reported above for Model B. As for the associated nucleosynthesis, similar amounts of unburned H (0.04, by mass) and $^4$He (0.1) are found at the end of the fourth burst (see Table \ref{tablenuc}). The mean, mass-averaged metallicity at the end of this fourth bursting episode is $Z \sim 0.86$ for this model. The most abundant species in the envelope are $^{60}$Ni (0.29), $^{64}$Zn (0.17), $^{4}$He (0.1), $^{32}$S (0.07), and $^{56}$Ni (0.06). A mass-averaged $^{12}$C yield of $\sim 0.02$ is also obtained at the end of the fourth burst, while the nucleosynthetic endpoint corresponds now to $^{103}$Ag, slightly above the endpoints reported for Models A \& B. Indeed, as shown in Table \ref{tablenuc}, there is a trend when comparing the nucleosynthesis associated with Models A, B, \& C: a small increase in the final abundances of intermediate-mass and heavy elements, likely driven by the specific choice of $^{15}$O($\alpha$,$\gamma)^{19}$Ne rate, since a larger adopted value for the rate will translate into a more effective breakout from the CNO cycles.

\section{Conclusion}

In summary, we have evaluated the available experimental data on the states in $^{19}$Ne that serve as resonances in the $^{15}$O($\alpha$,$\gamma)^{19}$Ne reaction in Type I x-ray bursts.  Using the likelihood distributions inferred from $\alpha$ decay branching ratio and lifetime measurements, we calculated the thermally averaged $^{15}$O($\alpha$,$\gamma)^{19}$Ne reaction rate via the Monte Carlo method, determining the median and 99.73\% confidence level upper and lower limits on the reaction rate. At this confidence level, the uncertainty in the rate is very large, exceeding a factor of 1000 at some temperatures. Given this large uncertainty, the central value of the rate, which does not differ from the recommended rates of \citet{cf88} or \citet{fisker06} by more than a factor of three at these temperatures, is not particularly meaningful and the whole range of reaction rates allowed by experiment must be considered.

Nevertheless, with the exception of the weak nucleosynthetic trend outlined above, we conclude that variation of the $^{15}$O($\alpha$,$\gamma)^{19}$Ne reaction rate within the present 99.73\% confidence level allowed range has no substantial effect on either the burst properties or the accompanying nucleosynthesis in our models of Type I x-ray bursts. In striking contrast to \citet{fisker06}, we do not find that a small value of the $^{15}$O($\alpha$,$\gamma)^{19}$Ne reaction rate at the lower limit allowed by experiment leads to steady state burning but rather to marginally more energetic bursts than are predicted with the recommended value of the reaction rate. Further astrophysical model calculations are required to resolve the apparent discrepancy between different sensitivity studies concerning the importance of the $^{15}$O($\alpha$,$\gamma)^{19}$Ne reaction in models of Type I x-ray bursts.

\section{Acknowledgements}
B.D. acknowledges support from the Natural Sciences and Engineering Research Council of Canada. TRIUMF receives federal funding via a contribution agreement through the National Research Council of Canada. R.H.C. would like to acknowledge helpful discussions with Manoel Couder and Wanpeng Tan. The work of R.H.C. was supported by the U.S. National Science Foundation Grant PHY-08-22648 (JINA). This work has been partially supported by the Spanish MICINN grants AYA2010-15685 and EUI2009-04167, by the E.U. FEDER funds, and by the ESF EUROCORES Program EuroGENESIS.

\bibliography{ne19}

\end{document}